\documentclass[conference]{IEEEtran} %journal
\ifCLASSINFOpdf
\else
\fi
\usepackage{cite}
\usepackage{amsmath}
\usepackage{amssymb}
\usepackage{amsfonts}
\usepackage{psfrag}
\usepackage{epsfig} 
\usepackage{algorithm}
\usepackage{algorithmic}
\usepackage{amsthm}

\newcommand{\B}[1]{\boldsymbol{#1}}
\begin{document}
\title{The information and wave-theoretic limits of analog beamforming} %CDI


\author{\IEEEauthorblockN{Amine Mezghani and Robert W. Heath, Jr.}
\IEEEauthorblockA{Wireless Networking and Communications Group\\
Department of ECE, The University of Texas at Austin\\
Austin, TX 78712, USA\\ 
Email: \{amine.mezghani, rheath\}@utexas.edu} } 
\maketitle

\begin{abstract}
The performance of broadband millimeter-wave (mmWave) RF architectures, is generally determined by mathematical concepts such as the Shannon capacity. These systems have also to obey physical laws such as the conservation of energy and the propagation laws. Taking the physical and hardware limitations into account is  crucial for characterizing the actual performance of mmWave systems under certain architecture such as analog beamforming. In this context, we consider a broadband frequency dependent array model
 that explicitly includes incremental time shifts instead of phase shifts between the individual antennas and incorporates a physically defined radiated power. 
%As a consequence of this model, the conventional separation of the waveform generation and beamforming problems becomes sub-optimal.
 As a consequence of this model, we present a novel joint approach for designing the optimal waveform and beamforming vector for analog beamforming.  Our results show that, for sufficiently large array size, the achievable rate  is mainly limited by the fundamental trade-off between the analog beamforming gain and signal bandwidth. 
 % by the nature of analog beamforming rather than the actual number of antennas.
\end{abstract}


% \fi
\IEEEpeerreviewmaketitle

\section{Introduction}  
The millimeter wave (mmWave) band offers a much higher available bandwidth which is a key ingredient for enabling high data rates in next-generation mobile cellular systems \cite{Rappaport_14,Boccardi_14,Bai_14,Swindlehurst_14}. Due to the required high number of antennas \cite{Swindlehurst_14,Marzetta_10,Larsson_14}  to compensate for the low SNR per antenna element, this technology creates several challenges at the same time, particularly in terms of hardware complexity. Analog processing based on phase shifters and the more general hybrid architecture \cite{Rappaport_14} are widely considered techniques for reducing the hardware complexity. The objective of having large bandwidth and large antenna gain simultaneously requires a careful performance analysis that is consistent with the physical limitations.  In fact, as an important part of such communication system is governed by electromagnetic theory and by antenna theory, a pure mathematical treatment of communication systems without consistent link to physical quantities such as radiated power might be questionable.  

The importance of using wave-theoretic or circuit based models for antennas arrays has been investigated in some previous and recent works dealing mainly with the narrowband case \cite{Loyka_2005,Ivrlac_10,Ivrlac_1014,Laas_17}. Thereby, the impact of antenna spacing and coupling on the information theoretic results of multiple antenna systems has been studied with a circuit based definition of power in \cite{Ivrlac_10,Ivrlac_1014,Laas_17}.  An insightful and general connection between electromagnetic wave theory and information theory in terms of number of degrees of freedom for the signal waveform is provided in \cite{Franceschetti_2015,Franceschetti}. In State-of-the art research on the performance of  mmWave systems with analog beamforming, however, generally lacks methodologies for deriving information theoretic results in accordance to wave-theoretic aspects and under certain  hardware restrictions. In fact, it is known in the classical antenna theory that there is a fundamental trade-off between the maximal achievable gain and achievable bandwidth \cite{Harrington,Hansen}. These classical results, however, do not consider the effect of analog processing and do not provide a simple information-theoretic interpretation.   

In this paper, we study the fundamental limits of analog transmit beamforming that is common across frequency given a certain radiated power. To this end, we adopt a broadband array model including delay shifts between the antenna elements \cite{Brady_2015}. We define the radiated power by the surface integral of the squared field over a sphere enclosing the antenna array \cite{Hansen}. The total radiated power plays an important role for the design of such mmWave systems not only from energy efficiency point of view but also due to regulatory restrictions and interference issues. As a consequence, the spatial precoding and the temporal waveform generation are coupled and cannot be considered independently. Therefore, we formulate a rate maximization problem under a certain total radiated power constraint assuming analog beamforming under single-path channel condition. The optimization parameters are jointly the spatial beamforming vector and the spectral shape.  The combined wave-theoretic and information theoretic analysis reveals a fundamental directivity-bandwidth trade-off limiting the achievable rate with analog beamforming. It shows that, for sufficiently large array size, the maximal achievable capacity is mainly limited by the frequency independent analog beamforming rather than the actual number of antennas. This finding constitutes a clear indication towards maintaining a separate RF chain for each antenna to fully exploit the potential of very large antenna arrays.  \\

\section{System and channel model}
\label{sec:system}

%%%%%%%
\begin{figure}[h]
\centerline{\includegraphics[width=8cm]{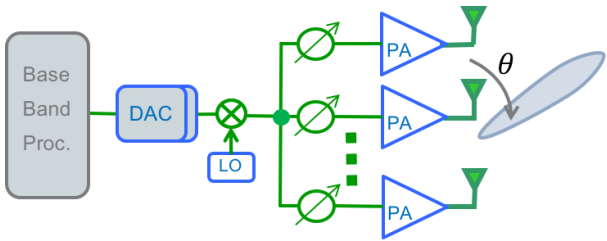}}
\caption{Analog beamforming architecture at the transmitter.}
\label{system_model}
\end{figure}
%%%%%
%%%%%%%%%   
\begin{figure}[h]
\psfrag{a}[c][c]{(a)}
\psfrag{b}[c][c]{(b)}
\centerline{\includegraphics[width=8cm]{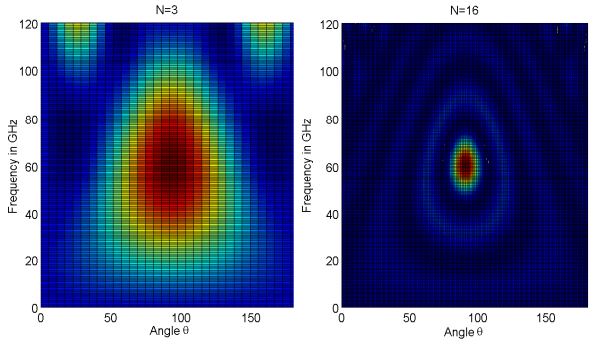}}
\caption{Radiation intensity as function of the azimuth angle and frequency for a circular array, $N\in\{3,16\}$. Smaller beamwidth implies smaller bandwidth.}
\label{radiation_pattern}
\end{figure}

We consider a single-user mmWave system, where a transmitter and a receiver are communicating via a single stream using analog beamforming.
% This transmission strategy is particularly relevant for this frequency band to compensate the low per-antenna SNR and due to the sparse nature of channels.
 We focus in this paper on the  transmitter side. We assume that the receiver perfectly selects its beam in the dominant line-of-sight (LOS) or non-LOS (NLOS)  direction. The beamforming gain at the receiver is then simply considered as part of the channel. The beamforming at the transmitter as illustrated in Fig.~\ref{system_model} is performed with $N$ antenna elements in the analog domain subject to a certain total radiated power constraint while the temporal signal shaping is done in the digital domain. Due to the angular selectivity of the receiver, the resulting channel transfer function including the receive beamforming is approximately described in the frequency domain by single dominant LOS or NLOS path from the point %of view of the transmitter
\begin{equation}
\begin{aligned}
  \B{h}(f)=  \alpha_{\rm c} \B{a}(\theta_{\rm c},\varphi_{\rm c},f),
\end{aligned}
\label{h(f)}
\end{equation}
where $\alpha_{\rm c}$ is the path coefficient (including path phase and strength), $\B{a}(\theta_{\rm c},\varphi_{\rm c},\omega)$ is the far-field array impulse response for the azimuth and elevation angles-of-departure (AoD)  $\theta_{\rm c}$ and $\varphi_{\rm c}$ in a spherical coordinate system as a function of the frequency $f$. The single-path assumption is made for simplicity only, and is not essential to our purpose of studying the limitations of analog beamforming. Further, we adopt a frequency dependent array response, which we refer to as the broadband array model. Note that the terminology ``narrowband"' or ``broadband" refers here to the frequency behavior of the antenna response and not to the propagation channel, which is assumed to be flat. Even when the individual antenna response is frequency flat, the frequency dependency of the array response might still result from the group delays between these elements. This fact is often neglected in the literature, where only  phase shifts are taken into account to describe a frequency flat array response. The broadband array model is however more appropriate in the context of mmWave systems as the array size might become electrically larger than the total group delay. In other words, denoting the signal bandwidth by $B$ and the maximal array size by $D$,  the narrowband condition $\frac{B \cdot D}{c} \ll 1$ ($c$: speed of light) is generally unjustified in mmWave systems with large array size and bandwidth of several GHz. \\

For a uniform linear array (ULA) of  hypothetical isotropic antennas with element spacing $d$ in wavelengths at the center frequency $f_{\rm c}$ and dimension $N$, the  broadband frequency response in the passband assuming all the frequencies propagate with the same speed is \cite{Brady_2015}
\begin{equation}
\begin{aligned}
&\B{a}(f,\theta)^{\rm T}\!=\!\!\left[1,\cdots\!, {\rm e}^{-{\rm j}{2 \pi}d \cos(\theta)n \frac{f}{f_{\rm c}}},\cdots,  {\rm e}^{-{\rm j}{2 \pi}d \cos(\theta) (N-1) \frac{f}{f_{\rm c}}} \right],
\end{aligned}
\label{array_res}
\end{equation}
where $f_{\rm c}$ is the center frequency of the occupied band $[f_{\rm min},f_{\rm max}]=[f_{\rm c}-B/2,f_{\rm c}+B/2]$, i.e., $f_{\rm c}=(f_{\rm min}+f_{\rm max})/2$. The term $d \cos(\theta)\frac{f}{f_{\rm c}}$ accounts for the time shift between adjacent antennas in the frequency domain and cannot be approximated by just a phase shift (with $f/f_{\rm c} \approx 1$) if $N\cdot (f_{\rm max}-f_{\rm min})/f_{\rm c} \not\ll 1$ as explained earlier. 

In analog beamforming, the transmitter applies a pulse shaping filter $p_0(f)$ in the digital domain and a frequency independent beamforming vector $\B{b}_0$ in the analog domain to the data signal.  Both yield the following structured spatial-temporal processing vector % to shape the signal in space and in time
%,  described in the frequency domain by the structured vector 
\begin{equation}
 \B{b}(f) = \B{b}_0 \cdot p_0(f).    
\end{equation}
%where $p_0(f)$ is the pulse shaping filter implemented in the digital filter, while $\B{b}_0$ is a frequency independent beamforming vector implemented in the analog domain.
 In other words, the analog precoding part is common over the entire bandwidth and cannot be adapted over the frequency. The restriction of the analog beamforming vector $\B{b}_0$  to be frequency independent is  for practical reasons and constitutes the major constraint in terms of performance as shown later.
%\footnote{An unstructured, i.e., unrestricted $\B{b}(f)$ is practically only possible if it is fully implemented in the digital domain.} 
 In addition to the frequency independence, the vector $\B{b}_0$ is usually subject to a constant modulus constraint due to the implementation using phase shifters. As we are interested in information and wave theoretical performance limits, this design constraint is not taken into account. \\
 
 Considering a single-carrier system with the channel vector from (\ref{h(f)}),  then the received signal in the frequency domain can be described as 
 \begin{equation}
 \tilde{y}(f) = \alpha_{\rm c} \B{a}(f,\theta_{\rm c},\varphi_{\rm c})^{\rm T}\B{b}(f) \tilde{x}(f)  +\tilde{z}(f),    
\end{equation}
with the information signal $\tilde{x}(f) $ having unit power spectral density and the noise $\tilde{z}(f)$ having the constant power spectral density $N_0$.
 The state of the art design of $\B{b}(f)$ has mainly evolved from the standard SISO approach, where the waveform generation through $p_0(f)$ and the spatial beamforming through $\B{b}_0$ are considered separately. In particular, $\B{b}_0$ is commonly chosen as the conjugate of the array response evaluated at the center frequency and the desired angular direction, i.e., $\B{b}_0 \propto \B{a}(f_{\rm c},\theta_{\rm c},\varphi_{\rm c})^*$. This method might be not optimal for broadband large antenna arrays due to frequency selective nature of the antenna array that leads to a coupled temporal and spatial behavior and a trade-off between bandwidth and antenna gain. As example, Fig.~\ref{radiation_pattern} shows the resulting total response of a circular array and its corresponding analog beamformer, i.e., the radiation pattern,  $|\B{a}(f,\theta)^{\rm T}\B{b}_0|^2$ designed at 60~GHz and $\theta_{\rm c}=90^{\circ}$ for sizes $N=3$ and $N=16$. We observe that the beamwidth and also bandwidth decrease simultaneously  with the number of antenna, in accordance to classical results from antenna theory.\\
 
  Another important physical quantities is the radiated power. The total radiated power plays an important role for the design of such mmWave systems not only from energy efficiency point of view, but also due to regulatory restrictions. Additionally, the radiated power at these frequencies is also limited compared to the sub-6~GHz frequencies because the implementation of efficient power amplifiers is quite challenging and costly at mmWave\footnote{Other radiation properties such as the EIRP are also restricted by regulation, which might also limit the maximal authorized antenna gain. This will not be taken into  account as we are interested in the physical limitations.}.  Due to conservation of energy, the radiated power is defined by the surface integral of the radiation intensity $|\B{a}(f,\theta,\varphi)^{\rm T}\B{b}_0p_0(f)|^2$ over a sphere enclosing the antenna array in the far field \cite{Balanis}
  
\begin{equation}
\begin{aligned}
 \int\limits_{f_{\rm min}}^{f_{\rm max}}  \frac{1}{4 \pi}\int\limits_0^{\pi}  \int\limits_0^{2\pi} \left|  \B{a}(f,\theta,\varphi)^{\rm T} \B{b} (f) \right|^2 \sin \theta ~ {\rm d}  \varphi ~ {\rm d}\theta ~ {\rm d} f \leq P_{\rm R}.
\end{aligned}
\label{Power_R}
\end{equation}  

A very common, but physically not necessarily consistent, definition of radiated power is based on the squared norm of the beamforming vector $\int \|\B{b}(f)\|^2 {\rm d}f$. This is equivalent to the physical definition in (\ref{Power_R}) only for the narrowband case with exactly half-wavelength antenna spacing \cite{Ivrlac_10}. 
 
Based on the above facts and considerations, we formulate in the next section the joint digital waveform and analog beamforming optimization in terms of achievable rate. 
%%%%%%%%%%%%%%%%%%%%%%%
%\begin {table}[thp]%
%\caption {Regulatory power limitation for the 60~GHz band \cite{Yong_11}}
%\label{power_lim}\centering %
%\begin{tabular}{|c|c|}
%\hline %
%Region & Power limitation \\
%\hline  %
%\hline
%USA/Canada &      10mW    \\\hline %
%Most other countries &    500mW     \\\hline %
%\end {tabular}
%\end {table}
%%%%%%%%%%%%%%%%%%%%%%%%%

\section{Achievable rate maximization under analog beamforming}
As a consequence of the coupling between the temporal and angular response in the broadband array model (\ref{array_res}), the goals of concentrating the signal in space (beamforming) and frequency (pulse shaping) should be considered jointly. The joint spatio-temporal spectral confinement is essential to characterize the actual achievable rate of the analog hardware architectures. Therefore, we formulate the following rate maximization problem under a certain total radiated power constraint assuming analog beamforming under the single-path transmission assumption:  

\begin{equation}
\begin{aligned}
&\max\limits_{\B{b}(f)=\B{b}_0p_0(f)}  \int\limits_{f_{\rm min}}^{f_{\rm max}} \log_2 \left(1+ \frac{1}{N_0}   \left| \alpha_{\rm c} \B{a}(f,\theta_{\rm c},\varphi_{\rm c})^{\rm T} \B{b}(f) \right|^2 \right) {\rm d} f   \\
& {\rm s.t.} ~ \int\limits_{f_{\rm min}}^{f_{\rm max}}  \frac{1}{4 \pi}\int\limits_0^{\pi}  \int\limits_0^{2\pi} \left|  \B{a}(f,\theta,\varphi)^{\rm T} \B{b} (f) \right|^2 \sin \theta ~{\rm d}  \varphi ~ {\rm d}\theta  ~ {\rm d} f \leq P_{\rm R}.
\end{aligned}
\label{optimization}
\end{equation} 
The optimization parameters are the spatial beamforming vector $\B{b}_0$ and the shaping filter $p_0(f)$. In the following, we restrict the analysis to the ULA case in (\ref{array_res}) and  we reformulate the problem in terms of angular-temporal spectrum. Particularly, we exploit the Vandermonde structure of the array response in (\ref{array_res}) to interpret the quantity $\B{a}(f,\theta_{\rm c})^{\rm T} \B{b}_0$ as the discrete Fourier transform (DFT) transform of the  vector elements in $\B{b}_0$.   In other words, we define the power spectrum density  $S_0( f)$ after the digital processing and the angular spectrum $G(\cos \theta_{\rm c} \cdot f)$ representing the analog processing part, using the substitutions
\begin{equation}
\begin{aligned}
G(\cos \theta \cdot f) &=  | \B{a}(f,\theta)^{\rm T} \B{b}_0  |^2, \\
S_0(f) & = | p_0(f)|^2,
\end{aligned}
\end{equation} 
%where $G(\cos \theta_{\rm c} \cdot f)$ represents the angular spectrum as function of the frequency based on the analog processing, while $S_0( f)$  is the power spectrum %density after the digital processing.
 Further, we assume an infinite number of antennas, as we are interested in the performance limits. Having unlimited number of antennas with half-wavelength spacing $d=1/2$, we can relax the angular spectral form  $G(\cdot)$ to be arbitrarily, but periodic with period $2f_{\rm c}$ (and satisfying the Dirichlet Fourier series conditions). Thus, we can obtain the asymptotic and simplified formulation with infinite array size
\begin{equation}
\begin{aligned}
\max\limits_{\B{b}(f)}  \int\limits_{f_{\rm min}}^{f_{\rm max}} \log_2 \left(1+ \frac{1}{N_0} G(\cos \theta_{\rm c} \cdot f) S_0( f)    \right) {\rm d} f \quad {\rm s.t.}    \\
 \int\limits_{f_{\rm min}}^{f_{\rm max}}  \frac{1}{2}\int\limits_0^{\pi}  G(\cos \theta \cdot f) S_0( f) \sin \theta  ~ {\rm d}\theta ~ {\rm d} f \leq P_{\rm R}, \\
 G(\cos \theta \cdot f) \geq 0,  S_0( f) \geq 0,~ \forall f, ~ \forall \theta.
\end{aligned}
\label{rate_opt_prob}
\end{equation} 
The optimization problem (\ref{rate_opt_prob}) is non-convex due to the bilinear form $ G(\cos \theta \cdot f) S_0(f)$ and difficult to solve in general. We provide instead the optimal solution for $S_0( f)$ given $G(\cos \theta \cdot f)$ and vice-versa. We introduce first the Lagrangian function for the case $S_0( f)>0$ and $G(\cos \theta \cdot f)>0$
\begin{equation}
\begin{aligned}
   L(G(\cdot),S_0(\cdot),\mu) \!=\! \int\limits_{f_{\rm min}}^{f_{\rm max}} \! \log_2 \left(1+ \frac{G(\cos \theta_{\rm c} \cdot f) S_0( f)}{{N_0}}    \right) {\rm d} f   \\
 - \mu \left(\int\limits_{f_{\rm min}}^{f_{\rm max}}  \frac{1}{2}\int\limits_0^{\pi}  G(\cos \theta \cdot f) S_0( f) \sin \theta ~ {\rm d}\theta ~ {\rm d} f - P_{\rm R}\right),
\end{aligned}
\label{lagrangian}
\end{equation}
with the Lagrangian variable $\mu$. For fixed  $G(\cos \theta \cdot f)$, the capacity-achieving $S_0( f)$ obtained by the KKT conditions follows from the well-known water-filling power allocation strategy over the frequency \cite{gallager}
\begin{equation}
\!\!\!\!S_0(f)\!= \!\frac{N_0}{\alpha_{\rm c}} \! \left( \!\frac{1}{ \frac{\mu}{2}\int\limits_0^{\pi}  G(\cos \theta \cdot f)  \sin \theta  {\rm d}\theta  }  -   \frac{1}{G(\cos \theta_{\rm c} \cdot f)} \!\right)_{\!\!\!\!+}\!\!,
\label{Sopt}
\end{equation}
for $f_{\rm min} \leq f \leq f_{\rm max}$, where $\mu$ is determined by the maximum power constraint in (\ref{rate_opt_prob}) and $(a)_+=\max(a,0)$. \\

Next, we consider the reverse case with fixed $S_0(f)$ and optimized $G(\cos \theta \cdot f)$. To this end, we rewrite the Lagrangian function (\ref{lagrangian}) using the substitutions $\Omega=\cos \theta \cdot f$ and $u=\cos \theta$ in a different way
\begin{equation}
\begin{aligned}
   L(G(\cdot),S_0(\cdot),\mu) \!=\! \int\limits_{f_{\rm min}}^{f_{\rm max}} \! \log_2 \left(1+ \frac{G(\cos \theta_{\rm c} \cdot f) S_0( f)}{{N_0}}    \right) {\rm d} f  - \\
 \mu \left( \int\limits_{0}^{2f_{\rm c}}   \frac{G(\Omega)+G(2f_{\rm c}-\Omega) }{2} \!\! \int\limits_{{\rm min}(\frac{\Omega}{f_{\rm max}},1)}^{{\rm min}(\frac{\Omega}{f_{\rm min}},1)}  \frac{S_0(\frac{\Omega}{u})}{u} {\rm d}u {\rm d} \Omega - P_{\rm R} \right),
 \end{aligned}
 \label{lagrange_2}
\end{equation}
where we made use of the periodicity of the function $G(\Omega)$ and the symmetry of the cosine function. The KKT condition corresponding to the maximization with respect to $G(\cos \theta_{\rm c} \cdot f)$ is obtained from the differential of (\ref{lagrange_2}) as follows
\begin{equation}
\begin{aligned}
    \frac{\frac{\alpha_{\rm c}}{N_0} S_0(f)}{1+ \frac{\alpha_{\rm c}}{N_0} G(\cos  \theta_{\rm c}  \cdot f) S_0(f)} -  \frac{\mu}{2}   \int\limits_{{\rm min}(\frac{\cos  \theta_{\rm c}  \cdot f}{f_{\rm max}},1)}^{{\rm min}(\frac{\cos  \theta_{\rm c}  \cdot f}{f_{\rm min}},1)}  \frac{S_0(\frac{\cos  \theta_{\rm c}  \cdot f}{u})}{u} {\rm d}u   \\
    -  \frac{\mu}{2}   \int\limits_{{\rm min}(\frac{2f_{\rm c}-\cos  \theta_{\rm c}  \cdot f}{f_{\rm max}},1)}^{{\rm min}(\frac{2f_{\rm c}-\cos  \theta_{\rm c}  \cdot f}{f_{\rm min}},1)}  \frac{S_0(\frac{2f_{\rm c}-\cos  \theta_{\rm c}  \cdot f}{u})}{u} {\rm d}u \stackrel{}{=}0,
 \end{aligned}
  \label{KKT_G}
\end{equation}
which can be solved with respect to $G(\cos  \theta_{\rm c}  \cdot f)$ in closed form. In the following we consider the solution for some particular cases in terms of $\theta_{\rm c}$.
\subsection{Solution around broadside of the ULA}
\label{broadside}
If $\cos \theta \leq f_{\rm min}/ f_{\rm max}$, then $\cos \theta \cdot f_{\rm max} \leq f_{\rm min}$ and $2f_{\rm c}- \cos \theta \cdot f_{\rm max} \geq 2f_{\rm c} - f_{\rm min}=f_{\rm max}$. Therefore (\ref{KKT_G}) simplifies to 

\begin{equation}
\begin{aligned}
    \frac{\frac{\alpha_{\rm c}}{N_0} S_0(f)}{1+ \frac{\alpha_{\rm c}}{N_0} G(\cos  \theta_{\rm c}  \cdot f) S_0(f)} -  \frac{\mu}{2}   \int\limits_{\frac{\cos  \theta_{\rm c}  \cdot f}{f_{\rm max}}}^{\frac{\cos  \theta_{\rm c}  \cdot f}{f_{\rm min}}}  \frac{S_0(\frac{\cos  \theta_{\rm c}  \cdot f}{u})}{u} {\rm d}u\stackrel{}{=}0.
 \end{aligned}
  \label{KKT_BS}
\end{equation}
We obtain then the  optimal solution for $G(\cdot)$ given $S_0(\cdot)$
\begin{equation}
G(\cos \theta_{\rm c} \cdot f)= \frac{N_0}{\alpha_{\rm c}} \left( \! \left( \frac{\mu}{2} \! \int\limits_{\frac{\cos  \theta_{\rm c}  \cdot f}{f_{\rm max}}}^{\frac{\cos  \theta_{\rm c}  \cdot f}{f_{\rm min}}}  \frac{S_0(\frac{\cos  \theta_{\rm c}  \cdot f}{u})}{u} {\rm d}u    \right)^{\!\!-1} \!\!\! -   \frac{1}{S_0(f)} \right)_{\!+}\!\!.
\label{Gopt}
\end{equation}

For the particular case of constant spectrum $S_0(f)$ across the entire bandwidth $B$, we deduce the following preposition.
\newtheorem {preposition1}{Preposition}
\begin {preposition1}
\label{preposition_broadside}
If $\cos \theta_{\rm c} \leq f_{\rm min}/ f_{\rm max}$, then the following angular and temporal spectral shapes provide a local minimum or a saddle point for the maximization (\ref{rate_opt_prob})
\begin{align}
S_0(f)&=\frac{P_{\rm R}}{B},  \\
G(\cos \theta_{\rm c} \cdot f)&= \frac{1}{|\cos \theta_{\rm c}|\log\sqrt{\frac{f_{\rm max}}{f_{\rm min}}}}, \label{Gmax}
\end{align}
for $f_{\rm min} \leq f \leq f_{\rm max}$, and zero otherwise. In other words, a spatio-temporal shape $G(\cos \theta \cdot f)S_0(f)$ which is flat over the bandwidth $B=f_{\rm min} - f_{\rm max}$ and a certain frequency dependent beamwidth satisfying $\cos \theta_{\rm c} \cdot f_{\rm min}  \leq \cos \theta \cdot f \leq \cos \theta_{\rm c} \cdot f_{\rm max}$ is a potential optimal solution.
\end {preposition1}
\begin{proof}
Since flat (constant) $S_0(f)$ and $G(\cos \theta \cdot f)$ can be shown to satisfy simultaneously the solutions for the alternating maximization (\ref{Sopt}) and (\ref{Gopt}), they solves the joint KKT conditions and are therefore potential joint maximizers of the achievable rate. 
\end{proof}
Preposition~1 implies that the maximum antenna gain obtained with flat spectrum is, except for $\theta_{\rm c}=\pm \pi/2$ (broadside), finite regardless of the number of antennas and can maximally reach the value in  (\ref{Gmax}). As example, consider a base station antenna configuration with a given sector size of $\pm 60^\circ$ around the broadside operating in the 27.5-28.35~GHz  band (intended for 5G \cite{Boccardi_14}), then the ULA gain is given by
\begin{equation}
\begin{aligned}
G_{\rm max,ULA,28~GHz}= \frac{1}{|\cos 60^\circ |\log\sqrt{\frac{28.35}{27.5}}} \approx 21.2{\rm dB}.
 \end{aligned}
\end{equation}
Higher frequency bands with larger bandwidth, for instance at 60~GHz might be limited by even lower maximum flat gain. Deploying other antenna configurations such as planar array can, however,  improves this gain substantially.
%%%%
\subsection{Solution in the end-fire direction of the ULA} 
\label{endfire}
The end-fire direction $\theta_{\rm c}=0$ is a limiting case that produces the maximal delay between the antennas. We expect therefore a more severe trade-off between  antenna gain and bandwidth. In the narrowband case, however, it is known that the antenna gain might scale superlinearly with the number of antennas \cite{Schelkunoff,Ivrlac_10}.  This phenomenon called ``super-gain" occurs at element spacing smaller than half-wavelength and requires low-loss antennas and narrowband operation \cite{Ivrlac_2010}.  Here, we aim instead at analyzing the broadband case with half-wavelength antenna spacing. To this end, we assume a flat temporal spectrum  $S_0(f)=P_{\rm R}/B$ across the available bandwidth $B=f_{\rm max}-f_{\rm min}$ and solve (\ref{KKT_G}) for $\theta_{\rm c}=0$ in terms of $G(\cdot)$. The solution reads as
\begin{equation}
G(f)= \frac{BN_0}{\alpha_{\rm c} P_{\rm R}} \left(\frac{\mu}{\log\left(\frac{f_{\rm max}}{\sqrt{f(2f_{\rm c}-f)}}\right)}-1\right)_+,
\end{equation}
where $\mu$ is chosen to satisfy the radiated power constraint in (\ref{rate_opt_prob}). Hence, the resulting radiation pattern is not flat as in the previous case, and leads to the following achievable rate in bit/s 
\begin{equation}
R_{\rm end-fire}= \int\limits_{f_{\rm min}}^{f_{\rm max}} \left(   \log_2\mu -  \log_2 \log\left(\frac{f_{\rm max}}{\sqrt{f(2f_{\rm c}-f)}}\right) \right)_+ {\rm d} f.
\end{equation}
%Additionally, alternating the individual optimizers Eqs. () can provide a local optimum solution. 

In the following section, we consider some numerical examples to illustrate the behavior of the data rate for both cases and at different frequency bands.  

\section{Numerical example}
\begin{figure}[h]
\psfrag{a}[c][c]{(a)}
\psfrag{b}[c][c]{(b)}
\centerline{\includegraphics[width=9cm]{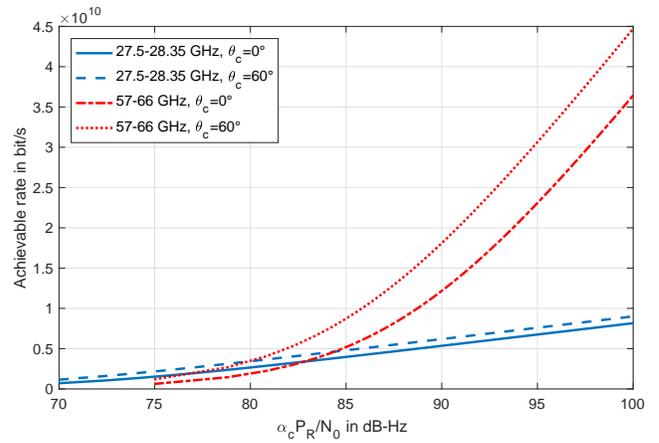}\vspace{-0.2cm}}
\caption{Achievable rate vs. the carrier-to-noise density ratio (C/N) with analog beamforming for the 28~GHz and 60~GHz bands. The 60~GHz band has lower achievable rate at small C/N despite the much larger bandwidth, which is due to the bandwidth-beamwidth trade-off.}
\label{num_example}
\end{figure}

\begin{figure}[h]
\psfrag{a}[c][c]{(a)}
\psfrag{b}[c][c]{(b)}
\centerline{\includegraphics[width=9cm]{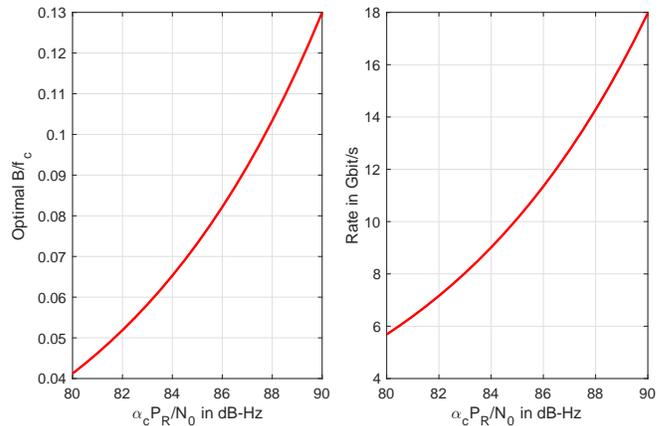}\vspace{-0.2cm}}
\caption{Optimal bandwidth and achievable rate for $f_{\rm c}=60$~GHz and $\theta_{\rm c}=60^{\circ}$ vs. the carrier-to-noise density ratio (C/N) with flat spectral. Large bandwidth is only meaningful for sufficiently high C/N.}
\label{num_example2}
\end{figure}

We apply our results from the previous section to the two widely-considered mmWave bands at 28~GHz with $27.5 {\rm ~GHz} \leq f \leq 28.35 {\rm ~GHz}$, and 60~GHz with $57 {\rm ~GHz} \leq f \leq 66 {\rm ~GHz}$. We choose two possible directions at $\theta_{\rm c}=60^{\circ}$ ($30^{\circ}$ apart from broadside) and $\theta_{\rm c}=60^{\circ}$ (end-fire). For $\theta_{\rm c}=60^{\circ}$, we have $\cos \theta_{\rm c} \leq f_{\rm min}/ f_{\rm max}$ for both bands and we can apply the results from Sub-section~\ref{broadside}, while for $\theta_{\rm c}=0^{\circ}$ we use the results from Sub-section~\ref{endfire}. The achievable rate with analog beamforming and infinite number of antennas is depicted in Fig.~\ref{num_example} versus the carrier-to-noise density ratio (C/N)  $\alpha_{\rm c} P_{\rm R}/N_0$. As expected, the achievable rate in the end-fire direction is lower than around the broadside. More interestingly, the 60~GHz band is more affected by the trade-off between bandwidth and beamwidth particularly in the low C/N regime and the larger bandwidth cannot be exploited efficiently. In fact, the 60~GHz band performs even worse than the 28~GHz when the entire available bandwidth is used at low C/N values. For this reason, we consider the optimization of the achievable rate based on the results from  Preposition~1 with respect to the bandwidth $B=f_{\rm max}-f_{\rm min}$ that should be used for the 60~GHz band, given $\theta_{\rm c}$ and $\alpha_{\rm c} P_{\rm R}$, i.e.,
\begin{equation}
\begin{aligned}
\max\limits_{B \leq 2f_{\rm c} \frac{1-\cos \theta_{\rm c}}{1+\cos \theta_{\rm c}}} R=B  \log_2 \left(1 + \frac{P_{\rm R}}{BN_0 |\cos \theta_{\rm c}|\log\sqrt{\frac{f_{\rm c}+B/2}{f_{\rm c} - B/2 }}} \right).  \nonumber % ~{\rm s.t.}~ |\cos \theta_{\rm c}| \leq \frac{f_{\rm c}-B/2}{f_{\rm c}+B/2 }
 \end{aligned}
\end{equation}
The results of this optimization are shown in Fig.~\ref{num_example2} for $\theta_{\rm c}=60^{\circ}$ and $f_{\rm c}=60$~GHz. The figure illustrates that the optimal bandwidth is sensitive to the C/N level and scales similarly to the rate. These observations apply for other mmWave frequency bands as well. 
%%%%%%%%%   
%%%%%%%
%%%%%%%%%   
%%%%%%%

%%%%%%%%%%%%%%%%%%%%%%%%%%%%%%%%%%%%%%%%%%%%%%%%%%%%%

\section{Conclusion}
We showed that analog beamforming with common coefficients across the frequency has a limited capacity regardless of the number of antennas. This limitation results from the fundamental trade-off between bandwidth and beamwidth of the resulting radiation pattern.  The analysis reveals that larger bandwidth is not necessary beneficial for the achievable rate due the reduced antenna gain attained by analog beamforming. Consequently, the joint design of temporal and spatial signal shape becomes a key for achieving the best trade-off.   As future work, we aim at considering hybrid precoding and other antenna configurations to mitigate this limitation.  
\section*{Acknowledgment}
This research was partially supported by the U.S. Department of Transportation through the Data-Supported Transportation Operations and Planning (D-STOP) Tier 1 University Transportation Center and a gift by Huawei.



%\bibliographystyle{IEEEtran}
\bibliographystyle{IEEEtran}   
\bibliography{references}{}

\end{document}